\documentclass[aps,prl,reprint,superscriptaddress,amsmath,amssymb]{revtex4-1}
\pdfoutput=1 

\usepackage{hyperref,graphicx}

\newcommand*{\bra}[1]{\langle #1 \vert}
\newcommand*{\ket}[1]{\vert #1 \rangle}

\newcommand*{\di}{\mathrm{d}}

\begin{document}

\title{Adiabatic Quantum Search in Open Systems}

\author{Dominik S. Wild}
\affiliation{Department of Physics, Harvard University, Cambridge, MA 02138, USA}

\author{Sarang Gopalakrishnan}
\affiliation{Department of Physics and Walter Burke Institute, California Institute of Technology, Pasadena, CA 91125, USA}
\affiliation{Department of Engineering Science and Physics, CUNY College of Staten Island, Staten Island, NY 10134, USA}
\author{Michael Knap}
\affiliation{Department of Physics, Walter Schottky Institute, and Institute for Advanced Study, Technical University of Munich, 85748 Garching, Germany}
\author{Norman Y. Yao}
\affiliation{Department of Physics, University of California, Berkeley, CA 94720, USA}
\author{Mikhail D. Lukin}
\affiliation{Department of Physics, Harvard University, Cambridge, MA 02138, USA}

\date{\today}

\begin{abstract}
  Adiabatic quantum algorithms represent a promising approach to universal quantum computation. In isolated systems, a key limitation to such algorithms is the presence of avoided level crossings, where gaps become extremely small. In open quantum systems, the fundamental robustness of adiabatic algorithms remains unresolved. Here, we study the dynamics near an avoided level crossing associated with the adiabatic quantum search algorithm, when the system is coupled to a generic environment. At zero temperature, we find that the algorithm remains scalable provided the noise spectral density of the environment decays sufficiently fast at low frequencies. By contrast, higher order scattering processes render the algorithm inefficient at any finite temperature regardless of the spectral density, implying that no quantum speedup can be achieved. Extensions and implications for other adiabatic quantum algorithms will be discussed. 
  \end{abstract}


\maketitle

The adiabatic theorem provides a powerful tool to characterize the evolution of a quantum system under a time-dependent Hamiltonian. It underlies theoretical concepts ranging from  Landau-Zener transitions~\cite{Shevchenko2010} to Berry phase accumulation and experimental techniques such as adiabatic passage~\cite{Bergmann1998}. Adiabatic evolution can also serve as a platform for quantum information processing~\cite{Farhi2000,Farhi2001,Das2008,Bapst2013,Santoro2006,Laumann2015}. This paradigm bears some resemblance to simulated annealing: computation proceeds via smoothly varying a parameter to hone in on a solution encoded in the ground state of a specific Hamiltonian. Thus, a generic adiabatic quantum computation (AQC) proceeds in three steps. A physical system is first prepared in the known ground state of a simple initial Hamiltonian. The Hamiltonian is then adiabatically transformed into the desired one. Finally, the state of the system is measured and, assuming adiabaticity, represents the solution to the encoded question.

\begin{figure}[t]
  \centering
  \includegraphics[width=\columnwidth]{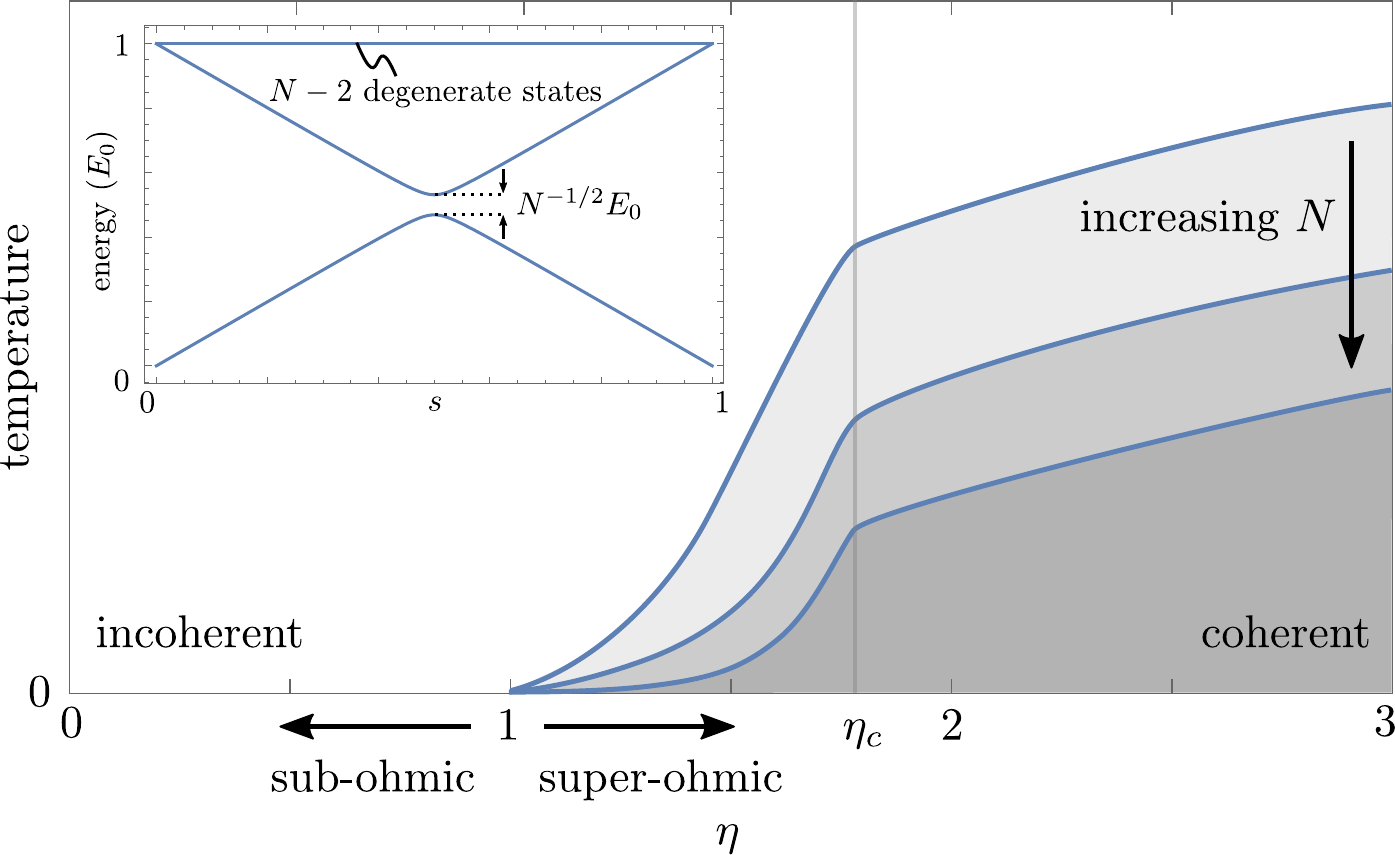}
  \caption{Qualitative dynamics of the adiabatic quantum search algorithm in an open system. The evolution of the system is coherent below the critical temperature $T^*$ (indicated by the solid curves) and a quantum speedup is available in this regime. The three curves correspond to different sizes $N$ of the search space. The parameter $\eta$ characterizes the noise spectral density at low frequency $\propto \omega^\eta$, with $\eta =1$ corresponding to an ohmic bath. The dependence of $T^*$ on $\eta$ changes qualitatively at $\eta_c$ due to scattering processes contributing significantly when $\eta > \eta_c$. The inset shows the spectrum of the AQS Hamiltonian for $N=256$.}
  \label{fig:phase_diagram}
\end{figure}

Nearly a decade ago, it was shown that  AQC and the canonical circuit model of quantum computation are equivalent in computational power~\cite{Siu2005,Aharonov2007,Mizel2007}. While the two models can provably solve the same problems, their physical implementation and thus their susceptibility to errors differ significantly. For instance, imperfections of individual gates will reduce the fidelity of a computation in the circuit model. In AQC, by contrast, errors may arise due to non-adiabatic transitions. Furthermore, AQC is affected by noise present in any realistic implementation. It has been suggested that AQC may be inherently robust against noise~\cite{Childs2001,Albash2015} and that the presence of an environment may even improve performance~\cite{Amin2008}. Adiabatic evolution is particularly susceptible to noise when the gap between the ground state and the excited states is small. A thorough understanding of the effect of noise on small gaps is therefore desirable. 
 In this Letter, we study the effect of an environment on the adiabatic quantum search (AQS) algorithm~\cite{Farhi2000,Roland2002}, the adiabatic equivalent of Grover's algorithm~\cite{Grover1997}. While the AQS algorithm in open systems has been the subject of numerous studies, a complete understanding of its scalability is missing~\cite{Shenvi2003,Aberg2005a,Aberg2005b,Roland2005,Ashhab2006,Tiersch2007,Amin2008,Perez2009,Avron2010,DeVega2010}. 
 
 Although the AQS algorithm involves a highly non-local Hamiltonian, we utilize it as a convenient example of an algorithm exhibiting a single avoided level crossing. In realistic systems with $k$-local interactions ($k \leq 2$ typically), small gaps often arise due to avoided level crossings between macroscopically distinct states. In this case, an environment that also acts locally is incapable of inducing transitions between the two states involved in the crossing, and it predominantly leads to dephasing. To this end, in our model for the AQS algorithm, the environment only couples to the dephasing channel. We show that under these assumptions, the problem of determining the scalability of the algorithm can be cast into an implementation-independent form, parametrized by the minimum gap at the avoided level crossing. Thus, we expect our conclusions to generalize beyond the AQS algorithm.

To understand the main result of our work, it is helpful to consider the different ways in which the environment influences the algorithm. One naively expects that a thermal bath will degrade performance whenever the temperature exceeds the smallest gap encountered during the computation. However, this is not necessarily the case if the number of thermally accessible states is small~\cite{Amin2008}. In the AQS algorithm, there exist two low-energy states, separated by a large gap from higher excited states. These two low-lying states undergo an avoided level crossing (see inset of Fig.~\ref{fig:phase_diagram}). It is thus natural to assume that the environment can thermally mix these two states but does not give rise to higher excitations. Thermalization may then reduce the success probability by at most 50\%, which can be compensated for by repeating the algorithm multiple times~\cite{Amin2008}.

\begin{table}
  \caption{Scaling of the critical temperature $T^*$ with the size of the search space $N$ for a given  coupling strength $\alpha$ between the system and the environment. The scaling of $T^*$ is evaluated separately for processes involving one and two bosons of the bath. For a sub-ohmic environment, the one-boson processes render the dynamics incoherent even at zero temperature such that two-boson processes are never relevant.
  \label{tab:critical_temp}}
  \begin{ruledtabular}
    \begin{tabular}{l l l}
      & single-boson processes & two-boson processes\\[.2cm]
      $\eta < 1$ & $\alpha T^* = 0$  & \\
      $\eta > 1$ &  $\alpha T^* = O(N^{(\eta - 2)/2})$ & $\alpha^{2/(2 \eta + 1)} T^* = O(N^{-1/(4\eta + 2)})$\\
    \end{tabular}
  \end{ruledtabular}
\end{table}

Apart from leading to thermalization, the environment also renormalizes the gap at the avoided crossing. The effect is best understood by appealing to an analogy with a double-well system. In this picture, the two low-energy states of the AQS algorithm are spanned by the ground states of two wells, which are detuned from each other by a bias $\varepsilon$ and connected by a tunneling rate $\Delta$. The avoided crossing occurs at zero bias ($s = 1/2$ in Fig.~\ref{fig:phase_diagram}), for which the energy gap is equal to the tunneling rate. As mentioned above, a local environment predominantly gives rise to dephasing between the wells, whereas environment-induced transitions from one well to another are negligible. This dephasing suppresses coherent tunneling, which in turn results in a decrease of the minimum gap. Equivalently, this mechanism may be viewed as a consequence of the quantum Zeno effect, where the environment tends to localize the system in one of the wells by gaining information about its current state~\cite{Misra1977}. Coherent tunneling may vanish entirely if the coupling to the environment is sufficiently strong. We refer to this as the incoherent regime, as opposed to the coherent regime, where tunneling persists. The terminology reflects the fact that coherent Rabi oscillations can, in principle, be observed in the coherent regime, whereas the oscillations are overdamped if the system is incoherent. Any potential quantum speedup is lost in the incoherent regime, as discussed in detail below. Conversely, a quantum speedup is always available in the coherent regime provided the gap retains the same scaling with problem size as in a closed system.

In order to identify the relevant regimes, we compare the tunneling rate with the coupling rate to the environment. At zero temperature, the coupling rate is given by the noise spectral density of the environment, $J(\omega)$, evaluated at the gap frequency. The noise spectral density is assumed to obey a power law at low frequencies, $J(\omega) \propto \omega^\eta$, where we distinguish between sub-ohmic ($\eta < 1$), ohmic ($\eta = 1$), and super-ohmic ($\eta > 1$) environments. For a sub-ohmic environment, the ratio $J(\Delta)/\Delta$ diverges in the limit $\Delta \to 0$, suggesting that the system is incoherent at the avoided level crossing for large search spaces. If the environment is super-ohmic, the same reasoning predicts that even large systems remain coherent. This simple argument is indeed correct at zero temperature, while at finite temperature, bosonic enhancement and two-boson processes lead to significant modifications. We demonstrate that even for a super-ohmic environment, a quantum speedup can only be achieved below a certain critical temperature, whose dependence on $\eta$ and the size of the search space is summarized in Tab.~\ref{tab:critical_temp} and Fig.~\ref{fig:phase_diagram}. Notably, the critical temperature decays as a power law with the size of the search space, such that the AQS algorithm offers no improvement over a classical algorithm for large search spaces at finite temperature. 

We now proceed with detailed calculations. The AQS algorithm in a closed system is described by the Hamiltonian %
$H(s) = E_0 ( 1 - s ) \left(\mathbb{I} - \ket{\psi_0}\bra{\psi_0}\right) + E_0 s \left( \mathbb{I} - \ket{m}\bra{m} \right)$,
where $E_0$ sets the energy scale of the system, $\ket{\psi_0} = \frac{1}{\sqrt{N}} \sum_{x=1}^{N} \ket{x}$ is an equal superposition of all states in the search space, and $\ket{m}$ denotes the marked element to be found. The parameter $s$ is increased monotonically from its initial value $s = 0$ to its final value $s = 1$. The Hamiltonian $H(s)$ can be exactly diagonalized in the two-level subspace spanned by $\ket{m}$ and $\ket{m_\perp} = \frac{1}{\sqrt{N-1}} \sum_{x\neq m} \ket{x}$, where 
\begin{equation}
  H(s) = \frac{E_0}{2} \mathbb{I} - \frac{1}{2} \left[ \varepsilon(s) \tau^z + \Delta(s) \tau^x \right].
\end{equation}
Here, $\tau^i$ are the Pauli matrices acting on $\{\ket{m}, \ket{m_\perp}\}$, $\varepsilon(s)/E_0 = 2 s - 1 + 2 ( 1 - s )/N$, and $\Delta(s)/E_0 = 2\frac{\sqrt{N-1}}{N} ( 1-s)$. The orthogonal subspace is degenerate with constant energy $E_0$ (see inset of Fig.~\ref{fig:phase_diagram}). The spectrum exhibits an avoided level crossing at $s=1/2$, where the gap is of order $O(N^{-1/2})$ for large $N$. As anticipated, the low-energy Hamiltonian is equivalent to one describing two wells connected by a tunneling rate $\Delta$ and detuned from each other by a bias $\varepsilon$. Classically, the computation time scales linearly with the size of the search space $N$, whereas both Grover's algorithm and the AQS algorithm achieve a quadratic quantum speedup, scaling as $O(N^{1/2})$. The latter scaling, set by the inverse of the minimum gap, is provably optimal~\cite{Roland2002}\footnote{The proof of optimality in reference~\cite{Roland2002} straightforwardly generalizes to open systems, assuming that the environment contains no information about the marked state.}. 

To specify the environment, we envision that the AQS Hamiltonian is implemented using $L$ qubits, where $N = 2^L$. Each qubit is coupled to an independent, bosonic bath. We assume throughout that the temperature $T \ll E_0/L$, which ensures that the dynamics of the system are restricted to the two lowest-lying levels. Under these conditions~\cite{SupMat}, the environment couples to the low-energy subspace through an effective interaction of the form
\begin{equation}
  V =  \tau^z \sum_k g_k ( b_k + b_k^\dagger ) + \tau^z \sum_{k, l} \frac{g_k g_l}{E} ( b_k + b_k^\dagger ) ( b_l + b_l^\dagger ) ,
  \label{eq:effective_hamiltonian}
\end{equation}
where $b_k$ and $b_k^\dagger$ are bosonic annihilation and creation operators, $g_k$ is a coupling strength, and $E$ an energy scale proportional to $E_0$. The first term in Eq.~(\ref{eq:effective_hamiltonian}) describes absorption or emission of a single boson, while the second term corresponds to two-boson processes, such as two-boson emission or boson scattering. Higher-order terms, which depend on specifics of the higher excited states, have been neglected since they do not affect our results qualitatively~\cite{SupMat}. We have also dropped terms that couple to $\tau^{x,y}$, representing environment induced transitions between $\ket{m}$ and $\ket{m_\perp}$, as they are strongly suppressed in the limit of large $N$~\cite{SupMat}. 

The bath is characterized by the noise spectral density $J(\omega) = \sum_k g_k^2 \delta(\omega - \omega_k)$, which follows a power law at low frequencies, $J(\omega) = \alpha \omega^\eta$. The parameter $\alpha$ sets the coupling strength to the environment. Our analysis is restricted to $\eta > 0$ because the effective two-level description breaks down otherwise~\cite{SupMat}. Furthermore, we assume that the weak-coupling condition $J(\omega) \ll E_0$ is satisfied for all $\omega$. We emphasize that coupling is only weak compared to the overall energy scale of the system but may be strong compared to the gap between the low-energy states.

In order to explore the coherence properties of the system, we employ a procedure known as adiabatic renormalization, which has been widely put to use in the context of the spin--boson model~\cite{Leggett1987}. The method is particularly powerful as it is valid even for non-perturbative and non-Markovian environments. Adiabatic renormalization proceeds by eliminating modes of the environment that are fast compared to the tunneling rate. To a good approximation these oscillators adiabatically follow the system thereby reducing the bare tunneling rate $\Delta$ to a renormalized tunneling rate $\tilde \Delta$. The case $\tilde \Delta = 0$ corresponds to the incoherent regime introduced above, while in the coherent regime $\tilde \Delta > 0$. To compute $\tilde \Delta$, we first determine the energy eigenstates in the absence of tunneling. For the moment, we only consider single-boson processes and limit ourselves to the region near the avoided crossing, where $\varepsilon(s) \approx 0$. The eigenstates are given by $\ket{ \tau, \mathbf{n}} = e^{- i \tau^z S_1} \ket{\tau} \prod_k \ket{n_k}$, where $S_1 = i\sum_k \frac{g_k}{\omega_k} (b_k - b_k^\dagger)$, $\tau = m, m_\perp$ (corresponding to $\tau^z = \pm 1$), and $n_k$ are the occupation numbers of the bosonic modes. Physically speaking, the system is dressed by oscillators, whose displacements depend on the state of the system. Oscillators with frequencies much greater than the tunneling rate will adjust to the state of the system almost instantaneously, while slower oscillators must be accounted for more carefully. We hence define the renormalized tunneling rate between the states $\ket{m, \mathbf{n}}$ and $\ket{m_\perp, \mathbf{n}}$ as $\tilde \Delta_\mathbf{n} = \Delta \, \bra{m, \mathbf{n}} \tau^x \ket{m_\perp, \mathbf{n}}'$, where the prime denotes that only oscillators with frequencies satisfying $\omega_k  > \Omega$ should be taken into account. Here, $\Omega$ is a low-frequency cutoff, which may be self-consistently determined as $\Omega = p \tilde \Delta_\mathbf{n}$. The exact value of $p$ is irrelevant in what follows, provided that $p \gg 1$. Due to the dependence of $\tilde \Delta_\mathbf{n}$ on the occupation numbers, it is only possible to define a unique renormalized tunneling rate at zero temperature. Nevertheless, we can define a typical rate $\tilde \Delta$ by taking a thermal expectation value, yielding
\begin{equation}
  \tilde \Delta = \Delta \exp \left[ - 2 \int_{\Omega}^\infty \di \omega \frac{J(\omega)}{\omega^2} \coth \frac{\omega}{2 T} \right].
  \label{eq:renormalized}
\end{equation}

We first consider the above expression at $T = 0$. For a super-ohmic environment, the integral in the exponent remains finite as $\Omega \to 0$. For large $N$, we may set $\Omega$ to zero to a very good approximation such that $\tilde \Delta$ is proportional to $\Delta$. If the environment is ohmic or sub-ohmic, the integral exhibits an infrared divergence. There exists a critical coupling strength $\alpha^* \propto \Delta^{1-\eta}$ such that $\tilde \Delta = 0$ for all $\alpha > \alpha^*$. For $\alpha < \alpha^*$, the renormalized tunneling rate remains finite~\cite{SupMat}. The critical coupling strength tends to zero as $N \to \infty$, showing that the dynamics are incoherent in the limit of large search spaces consistent with the discussion above.

The results at finite temperature can be obtained by very similar arguments. In short, one obtains that $\tilde \Delta$ is always finite and proportional to $\Delta$ for $\eta >2$, while for $1 < \eta \leq 2$ there exists a critical coupling strength of the form $\alpha^* \propto \Delta^{2-\eta} / T$, where we assumed that $T \gg \tilde \Delta$. If the coupling constant is fixed, the expression can be interpreted as an expression for a critical temperature 
\begin{equation}
  T^* \propto \frac{\Delta^{2-\eta}}{\alpha} = O(N^{(\eta - 2)/2}).
\end{equation} 
This is consistent provided $\eta > 1$. In the sub-ohmic regime, $T^*$ cannot be taken much greater than $\tilde \Delta$ and we find instead that the dynamics are always incoherent for a fixed $\alpha$ in the limit of large search spaces. At $\eta = 1$, the existence of a non-zero critical temperature depends on the value of $\alpha$. We note that these results, summarized in the first column of Tab.~\ref{tab:critical_temp}, are in agreement with previous work by Tiersch and Sch\"utzhold~\cite{Tiersch2007}.

Two-boson processes may be treated similarly although they affect the system in a qualitatively different manner~\cite{Kagan1989}. There are two kinds of two-boson processes: those in which a pair of bosons is absorbed or emitted, and those in which a boson is scattered between two modes. Conservation of energy requires that in two-boson emission/absorption processes both modes have energies $\alt \Delta$. By contrast, the scattering processes can involve pairs of modes with arbitrarily high energy, provided their energy difference is small. 
Crucially, the phase space for boson scattering is independent of $\Delta$ for large $N$ and remains non-zero as $\Delta \rightarrow 0$. The two-boson coupling strength at finite temperature is thus expected to be always large compared to $\Delta$ for large N.

To support this argument, we again perform adiabatic renormalization~\cite{SupMat}. We focus on super-ohmic environments since single-boson processes already prevent a quantum speedup in the sub-ohmic case. We further extend the weak coupling approximation to include bosonic enhancement, i.e., $J(\omega) (1 + N(\omega)) \ll E_0$ for all $\omega$, where $N(\omega)$ is the Bose--Einstein distribution. Under these assumptions, two-boson processes only weakly renormalize the tunneling rate at zero temperature and do not render the dynamics incoherent. If $T > 0$, there exists a critical coupling strength, which is given by $\alpha^* \propto E \Delta^{1/2}/T^{\eta+1/2}$, such that the dynamics are incoherent for any $\alpha > \alpha^*$. Clearly, $\alpha^*$ vanishes as $N \to \infty$ regardless of $\eta$. This is in stark contrast to the renormalization due to single-boson processes alone, where the system remains coherent if $\eta > 2$. At fixed coupling strength, we thus predict a critical temperature 
\begin{equation}
  T^* \propto \frac{\Delta^{1/(2\eta + 1)}}{\alpha^{2/(2 \eta + 1)}} = O( N^{-1/(4\eta + 2)} )
\end{equation}
for two-boson processes.

In addition to coherent tunneling, there exist incoherent transitions, during which the system exchanges energy with the environment and thermalizes. We argued above that in the case of the AQS algorithm, these processes merely give rise to constant overhead. In fact, thermalization may even improve the performance if it occurs sufficiently fast~\cite{Amin2008}. By letting the system thermalize, one can obtain the ground state with a probability of at least 50\% since only the lowest two energy states may be significantly populated. In order to exclude the possibility of a quantum speedup in the incoherent regime, it is therefore necessary to ensure that the thermalization rate decreases with system size at least as fast as $O(N^{-1})$. Indeed, the thermalization rate always scales as $O(N^{-1})$ in the incoherent regime~\cite{SupMat}.

In the coherent regime, the thermalization rate can exceed this scaling near the avoided level crossing. This is an intriguing result since it implies that quantum computation can proceed through thermalization alone. This may be accomplished, for instance, by initializing the system in its ground at $s = 0$ (large bias) before rapidly decreasing the bias to zero. The system is then left to thermalize before being measured in the computational basis. Repeating this procedure several times will yield the ground state with high probability. We note, however, that this approach does not lead to an improved scaling compared to adabatic evolution, which always offers a quantum speedup in the coherent regime.

\begin{figure}[t]
  \centering
  \includegraphics[width=\columnwidth]{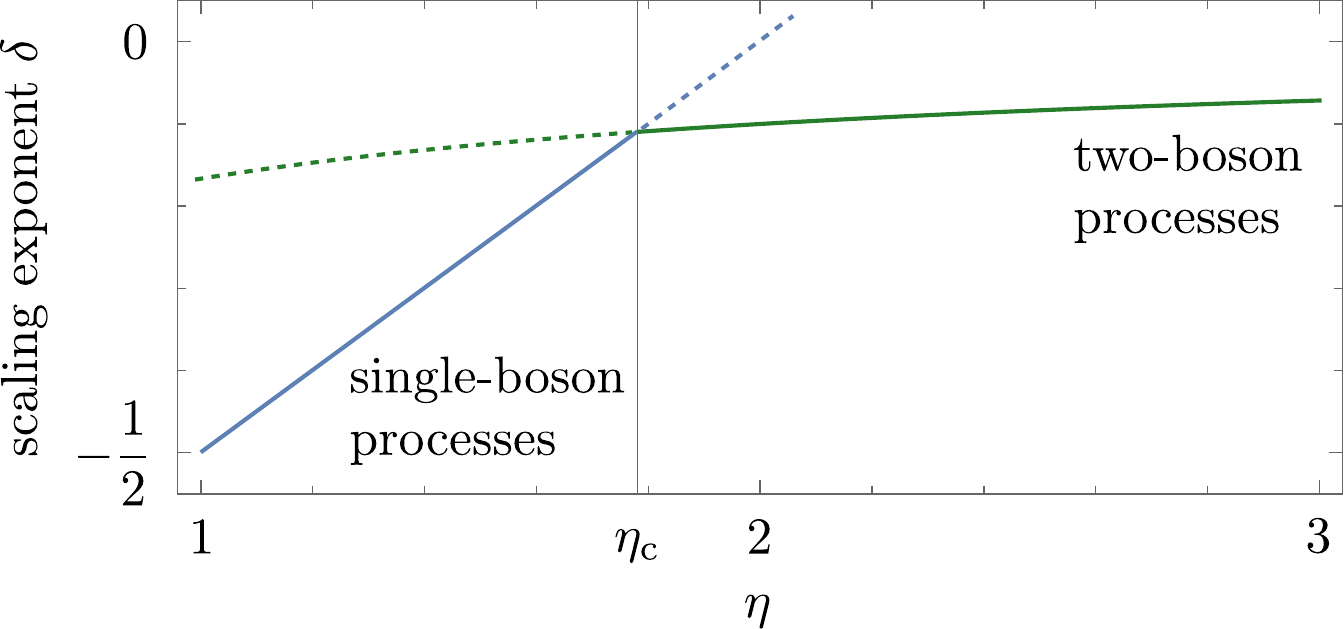}
  \caption{Dependence of the critical temperature $T^*$ on the search space size $N$. The critical temperature follows a power law $T^* = O(N^\delta)$. Above $T^*$, the system evolves incoherently, while below, quantum coherence is retained. The qualitative change at $\eta_c$ is due to competition between single and two-boson processes. For $\eta < 1$, the dynamics are incoherent even at zero temperature in the limit of large $N$.}
  \label{fig:critical_temp}
\end{figure}

We summarize our results by discussing the combined effect of single-boson and two-boson processes. In the parameter regime considered, the two processes decouple and their combined effect can be deduced from the results presented above~\cite{SupMat}. In particular, for the dynamics to be incoherent it is sufficient that one of the processes renormalizes the tunneling rate to zero. We thus conclude that the system is always incoherent at finite temperature in the limit of large $N$ and the algorithm does not provide a quantum speedup. We observe that the critical temperature associated with the coherent--incoherent transition scales differently for the two processes, see Fig.~\ref{fig:critical_temp}. Only the smaller critical coupling is physically significant; thus, two-boson processes dominate for $\eta > \eta_c$, and single-boson processes otherwise. At $\eta_c = (3+\sqrt{17})/4$ the critical temperatures scale identically and model-dependent pre-factors determine which process dominates.

Owing to the generic nature of the system--bath interaction discussed here, we expect that our results extend to a wide range of adiabatic algorithms involving avoided level crossings. The interaction Hamiltonian in Eq.~(\ref{eq:effective_hamiltonian}), involving only dephasing, arises naturally in such situations because small gaps generically correspond to macroscopically distinct states that are not connected by a local environment. The non-local interactions in the AQS algorithm lead to a spectrum in which the $N - 2$ states not involved in the level crossing are extensively separated in energy (i.e., their excitation gap is proportional to the full energy bandwidth of the system). A more realistic model with few-body interactions will instead have an intensive excitation gap. As long as the temperature is much lower than this excitation gap, our reduced model of the avoided crossing continues to apply, and so do our conclusions. Moreover, our findings should be broadly revelant for adiabatic quantum algorithms that involve many-body tunneling~\cite{Young2010,Denchev2015,Boixo2016}. For the AQS algorithm we were able to draw a direct correspondence between tunneling and speedup, whereas the general significance of tunneling in AQC algorithms is an open question. Future work may explore the applicability of our results to algorithms offering an exponential speedup, where the role of many-body tunneling is particularly unclear~\cite{Crosson2016}. Finally, our work highlights the need for quantum error correction to render AQC scalable at finite temperature~\cite{Bennett1996,Gottesman1997,Knill2000,Jordan2006,Lidar2008,Young2013,Pudenz2014,Bookatz2015}.

\begin{acknowledgments}
  We thank E.~Demler, V.~Oganesyan, J.~H.~Wilson, and L.~Zhou for insightful discussions. Financial support was provided by the NSF, the Center for Ultracold Atoms, and the NSSEFF program. SG is supported by the Walter Burke Institute. MK acknowledges support from the Technical University of Munich---Institute for Advanced Study, funded by the German Excellence Initiative and the European Union FP7 under grant agreement 291763. NYY is supported by the Miller Institute for Basic Research in Science.
\end{acknowledgments}

\bibliography{bibliography}

\end{document}


\title{Supplemental Materials for Adiabatic Quantum Search in Open Systems}
\maketitle

\onecolumngrid
\section{Derivation of the effective Hamiltonian\label{app:effective_hamiltonian}}
In this section, we derive the effective interaction Hamiltonian in Eq.~(2) of the main text from a microscopic model of the system--environment interaction. The Hamiltonian of the closed system was described in the main text and is given by
\begin{equation}
  H(s) = E_0 ( 1 - s ) \left( I - \ket{\psi_0}\bra{\psi_0} \right) + E_0 s \left( I - \ket{m} \bra{m} \right),
\end{equation}
where $\ket{m}$ is the marked state and $\ket{\psi_0} = \frac{1}{\sqrt{N}} \sum_{x=1}^N \ket{x}$. We assume that this Hamiltonian is implemented using $L$ qubits, such that the states $\ket{x}$ in the search space are represented by the $N = 2^L$ eigenstates of the Pauli operators $\left\{ \sigma^z_j \right\}$, with $j = 1, \dots, L$, acting on the individual qubits. We consider the situation where each qubit is coupled to an independent, bosonic bath, described by the generic interaction Hamiltonian
\begin{equation}
  V = \sum_{j=1}^{L} \sum_{\mu=x,y,z} \sigma_j^\mu \otimes \sum_k  g_{jk}^{\mu} \left( b_{jk}^{\mu} + b_{jk}^{\mu \dagger} \right).
  \label{eq:interaction}
\end{equation}
Here $b_{jk}^\mu$ are independent bosonic annihilation operators and $g_{jk}^\mu$ is the coupling strength to a particular mode. We further assume that the baths are identical such that the bath Hamiltonian is given by
\begin{equation}
  H_B = \sum_k \omega_k \sum_{j=1}^L \sum_{\mu = x, y, z} b_{jk}^{\mu \dagger} b_{jk}^{\mu}.
\end{equation}
We note that many of the present assumptions can be relaxed without affecting our results qualitatively. For instance, our calculation readily carries over to the situation where all qubits couple to the same environment provided the interaction remains local.

In a closed system, the excited states at energy $E_0$ are completely decoupled from the non-trivial subspace $\mathcal{S} = \mathrm{span}\left\{ \ket{m}, \ket{m_\perp} \right\}$. Although this is not the case in an open system, we can describe the dynamics of the subspace $\mathcal{S}$ near the avoided crossing by an effective Hamiltonian provided the steady-state population in the excited levels is negligible. In thermal equilibrium with a bath at temperature $T$, this gives rise to the condition $T \ll E_0/L$ near the avoided level crossing. Since $E_0$, the overall energy scale of the system, is an extensive quantity, this condition can be satisfied by a small but intensive temperature. In this limit, the effective Hamiltonian for the system and environment can be derived using the general formalism in reference~\cite{Cohen-Tannoudji2004}, yielding
\begin{align}
  H_\mathrm{eff} &= P H(s) P + H_B + P V  P + \frac{1}{2} \sum_{a,b,e} \left( \frac{1}{E_a - E_e} + \frac{1}{E_b - E_e} \right) \ket{a} \bra{a} V \ket{e} \bra{e} V  \ket{b} \bra{b},
  \label{eq:heff}
\end{align}
where $P$ is the projection operator onto $\mathcal{S}$. In the sum, the indices $a,b$ run over the eigenstates of $H(s)$ in $\mathcal{S}$ (low-energy states), while $e$ refers to the states in the orthogonal subspace (excited states). There are two contributions to the interaction of the environment with the low-energy states: a direct interaction and one that is mediated by the excited states through virtual processes. We have neglected higher order terms, which include couplings between different excited states. Such processes cannot be described in terms of the parameters of the avoided crossing alone and are therefore beyond the scope of our discussion. Furthermore, these processes do not affect our results qualitatively as discussed in more detail in section~\ref{app:two_boson}.

The quantum speedup in the closed system is enabled by tunneling near the avoided level crossing. We will therefore restrict ourselves to that region, which allows us to replace the energy differences in the denominator of the last term in Eq.~(\ref{eq:heff}) by $-E_0/2$, neglecting terms of order $\sqrt{\varepsilon^2 + \Delta^2}/E_0$. This drastically simplifies the expression to
\begin{align}
  H_\mathrm{eff} &\approx H_\mathcal{S} + H_\mathrm{B} + P V P - \frac{1}{2 E_0}  P V (I - P) V P ,
\end{align}
where
\begin{equation}
  H_\mathcal{S}(s) = P H(s) P = \frac{E_0}{2} - \frac{1}{2} \left[ \varepsilon(s) \tau^z + \Delta(s) \tau^x \right].
\end{equation}
Here $\tau^\mu$ are the Pauli matrices acting on $\mathcal{S}$. In the $\{\ket{m},\ket{m_\perp}\}$ basis, the projections of the Pauli operators acting on the physical qubits are given by (see the appendix of reference~\cite{Wan2009} for details)
\begin{equation}
  P \sigma_j^{x} P = \frac{1}{N-1}
  \begin{pmatrix}
    0 & \sqrt{N-1}\\
    \sqrt{N-1} & N-2
  \end{pmatrix} , \qquad
  P \sigma_j^{y} P  = \frac{s_j}{\sqrt{N-1}} 
  \begin{pmatrix}
    0 & -i\\
    i & 0
  \end{pmatrix} ,\qquad
  P \sigma_j^{z} P  = \frac{s_j}{N-1} 
  \begin{pmatrix}
    N-1 & 0\\
    0 & -1
  \end{pmatrix},
  \label{eq:pauli}
\end{equation}
where $s_j = \bra{m} \sigma_j^{(z)} \ket{m} = \pm 1$. We observe that the off-diagonal terms in these matrices are of order $O(N^{-1/2})$ so that coupling between $\ket{m}$ and $\ket{m_\perp}$ is suppressed in the limit of large $N$. This fact has the simple physical interpretation that $\ket{m}$ and $\ket{m_\perp}$ are macroscopically distinct, while the environment acts only locally. In the following, we neglect all terms of order $O(N^{-1/2})$ and below. 

The direct interaction of the environment with the low-energy subspace is hence given by
\begin{equation}
  V_1 = P V P \approx \frac{1}{2} \sum_{j=1}^L  \sum_k \left[ g_{jk}^x \left( I - \tau^z \right) \otimes \left( b_{jk}^x + b_{jk}^{x\dagger} \right) +  g_{jk}^z s_j \left( I + \tau^z \right) \otimes \left( b_{jk}^z + b_{jk}^{z\dagger} \right) \right],
\end{equation}
while the excited states mediate a two-boson interaction of the form
\begin{align}
  V_2 = - \frac{1}{2 E_0} P V (I -P) V \approx - \sum_{j=1}^L \sum_{k, l} &\left[ \frac{g_{jk}^x g_{jl}^x}{4 E_0} (I + \tau^z) \otimes \left( b_{jk}^x + b_{jk}^{x\dagger} \right) \left( b_{jl}^x + b_{jl}^{x\dagger} \right) + \frac{g_{jk}^y g_{jl}^y}{2 E_0} I \otimes  \left( b_{jk}^y + b_{jk}^{y\dagger} \right) \left( b_{jl}^y + b_{jl}^{y\dagger} \right) + \right. \nonumber\\
    & \left. + \frac{g_{jk}^z g_{jl}^z}{4 E_0} (I - \tau^z) \otimes  \left( b_{jk}^z + b_{jk}^{z\dagger} \right) \left( b_{jl}^z + b_{jl}^{z\dagger} \right) \right].
\end{align}
This follows from the observation that $P \sigma_i^\mu \sigma_j^\nu P \approx P \sigma_i^\mu P \sigma_j^\nu P$ when $i \neq j$. We note that the field $b_i^y$ only couples to the identity and therefore does not affect the dynamics of the system. In order to further simplify the expressions, let us focus on a single $x$ mode and simplify the notation by only retaining a single subscript for the mode label. The bath plus effective interaction Hamiltonian involving this mode takes the form
\begin{align}
  H_x = I &\otimes \left[ \sum_k \omega_k b_k^\dagger b_k + \sum_k \frac{g_k}{2} \left( b_k + b_k^\dagger \right) - \sum_{k,l} \frac{g_k g_l}{4 E_0} \left( b_k + b_k^\dagger \right) \left( b_l + b_l^\dagger \right) \right] - \nonumber\\
  - \tau^z &\otimes \left[ \sum_k \frac{g_k}{2} \left( b_k + b_k^\dagger \right) + \sum_{k,l} \frac{g_k g_l}{4 E_0} \left( b_k + b_k^\dagger \right) \left( b_l + b_l^\dagger \right) \right].
\end{align}
The terms coupling to the identity can be understood as the backaction of the system on the environment. We account for this effect by diagonalizing these terms, resulting in a set of renormalized bosonic operators. It is straightforward to check that the single-boson term can be accounted for by introducing the shifted operators
\begin{equation}
  c_k = b_k + \frac{1}{1 - a} \frac{g_k}{2 \omega_k}, 
\end{equation}
where
\begin{equation}
  a = \frac{1}{E_0} \sum_k \frac{g_k^2}{\omega_k} = \frac{1}{E_0} \int_0^\infty \di \omega \frac{J(\omega)}{\omega}.
\end{equation}
The existence of the integral requires that $J(\omega)$ decays sufficiently fast as $\omega \to \infty$. Furthermore, the power-law dependence $J(\omega) \propto \omega^\eta$ must satisfy $\eta > 0$. The Hamiltonian $H_x$ can hence be written as
\begin{align}
  H_x = I &\otimes \left[ \sum_k \omega_k c_k^\dagger c_k - \sum_{k,l} \frac{g_k g_l}{4 E_0} \left( c_k + c_k^\dagger \right) \left( c_l + c_l^\dagger \right) \right] - \nonumber\\
  - \tau^z &\otimes \left[ \frac{\tilde \varepsilon}{2} + \frac{1 - 2 a}{1 - a} \sum_k \frac{g_k}{2} \left( c_k + c_k^\dagger \right) + \sum_{k,l} \frac{g_k g_l}{4 E_0} \left( c_k + c_k^\dagger \right) \left( c_l + c_l^\dagger \right) \right],
\end{align}
where we introduced the environment-induced bias
\begin{equation}
  \tilde \varepsilon = \frac{E_0}{4} \frac{a(3a - 2)}{(1-a)^2}.
\end{equation}
This induced bias results in a shift of the avoided level crossing. Since the location of the avoided level crossing bears no significance, as long as it is known, we will drop the environment-induced bias in what follows.

The quadratic term coupling to the identity can be diagonalized perturbatively using the techniques outlined in section~\ref{app:diagonalization}. The result is that
\begin{equation}
  H_x = I \otimes \sum_k \tilde \omega_k d_k^\dagger d_k - \tau^z \otimes \left[ \sum_k \tilde g_k  \left( d_k + d_k^\dagger \right) + \sum_{k,l} \frac{\tilde g_k \tilde g_l}{\tilde E} \left( d_k + d_k^\dagger \right) \left( d_l + d_l^\dagger \right) \right],
  \label{eq:hx_final}
\end{equation}
where
\begin{gather}
  \tilde \omega_k = \omega_k - \frac{g_k^2}{2 E_0}, \\
  \tilde g_k = \frac{1 - 2 a}{1 - a} \left( 1 - \frac{g_k}{2 E_0} \sum_{l \neq k} \frac{g_l}{\omega_k - \omega_l} + \frac{g_k}{2 E_0} \sum_l \frac{g_l}{\omega_k + \omega_l} \right) \frac{g_k}{2}, \\
  \tilde E = \left( \frac{1-a}{1 - 2 a} \right)^2 E_0,
\end{gather}
and $d_k$ are new bosonic operators. They can be related to $c_k$ by
\begin{equation}
  d_k \approx c_k + \frac{g_k}{2 E_0} \sum_{l \neq k} \frac{g_l}{\omega_k - \omega_l} c_l - \frac{g_k}{2 E_0} \sum_{l} \frac{g_l}{\omega_k + \omega_l} c_l^\dagger.
\end{equation}
Before proceeding, it is worth verifying the validity of perturbation theory employed here. The correction of the energy eigenvalues is certainly small since $g_k^2$ is inversely proportional to the volume of the bath. It vanishes entirely in the thermodynamic limit and we will therefore neglect it below. In addition, we require that the correction to the bosonic operators be small, i.e.,
\begin{equation}
  \frac{g_k^2}{4 E_0^2} \left[ \sum_{l \neq k} \frac{g_l^2}{(\omega_k - \omega_l)^2} + \sum_l \frac{g_l^2}{(\omega_k + \omega_l)^2} \right] \ll 1.
  \label{eq:eigenvector_condition}
\end{equation}
Let us consider the first sum by re-writing it in terms of the noise spectral density $J(\omega)$. We introduce the mode spacing $\Delta \omega$ at frequency $\omega_k$ such that
\begin{equation}
  \frac{g_k^2}{4 E_0^2} \sum_{l \neq k} \frac{g_l^2}{\left( \omega_k - \omega_l \right)^2} = \frac{1}{4 E_0^2} \int_{\omega_k - \Delta \omega}^{\omega_k + \Delta \omega} \di \omega J(\omega) \left[ \int_0^{\omega_k - \Delta \omega} \di \omega' \frac{J(\omega')}{(\omega - \omega')^2} + \int_{\omega_k + \Delta \omega}^{\infty} \di \omega' \frac{J(\omega')}{(\omega - \omega')^2} \right].
\end{equation}
We are interested in the continuum limit $\Delta \omega \to 0$, which yields
\begin{equation}
  \frac{g_k^2}{4 E_0^2} \sum_{l \neq k} \frac{g_l^2}{\left( \omega_k - \omega_l \right)^2} = \frac{J(\omega_k)}{2 E_0^2} \lim_{\Delta \omega \to 0} \Delta \omega \left[ \int_0^{\omega_k - \Delta \omega} \di \omega' \frac{J(\omega')}{(\omega_k - \omega')^2} + \int_{\omega_k + \Delta \omega}^{\infty} \di \omega' \frac{J(\omega')}{(\omega_k - \omega')^2} \right] = \frac{J(\omega_k)^2}{E_0^2},
\end{equation}
where we employed L'H\^opital's rule to evaluate the limit. This shows that the first sum in Eq.~(\ref{eq:eigenvector_condition}) is small as long as the weak coupling limit $J(\omega) \ll E_0$ is satisfied. The second sum can be bounded from above by
\begin{align}
  \frac{g_k^2}{4 E_0^2} \sum_l \frac{g_l^2}{(\omega_k + \omega_l)^2} &\leq \frac{g_k^2}{4 E_0^2} \sum_{l \neq k} \frac{g_l^2}{\left( \omega_k - \omega_l \right)^2} + \frac{g_k^4}{16 E_0^2 \omega_k^2}\nonumber\\
  &\leq \frac{J(\omega_k)^2}{E_0^2} + \lim_{\Delta \omega \to 0} \Delta \omega^2 \frac{J(\omega_k)^2}{16 E_0^2 \omega_k^2}.
\end{align}
The last term is again small in the weak coupling regime since we can always take $\omega_k \geq \Delta \omega$ without modifying the spectrum of the bath significantly. Hence, perturbation theory is valid in the weak coupling regime.

The Hamiltonian in Eq.~(\ref{eq:hx_final}) indeed has the form of the effective Hamiltonian introduced in the main text. There will be a similar contribution for each qubit and polarization of the bath modes. Since these contributions all commute, we expect the dynamics to be well described by a single contribution, with the only modification being that $J(\omega)$ should multiplied by the number of channels that couple to the environment.

\section{Adiabatic renormalization\label{app:adiabatic}}
We argued in the main text that it is possible to treat the fast oscillators of the environment by introducing a renormalized tunneling rate between the two well states. More specifically, we consider the system and bath Hamiltonian
\begin{equation}
  H = - \frac{1}{2} \left( \varepsilon \tau^z + \Delta \tau^x \right)  + \sum_k \omega_k b_k^\dagger b_k + \tau^z \otimes \left[ \sum_k g_k \left( b_k + b_k^\dagger \right) + \sum_{k,l} \frac{g_k g_l}{E} \left(b_k + b_k^\dagger \right) \left( b_l + b_l^\dagger \right)  \right]
  \label{eq:hamiltonian}
\end{equation}
in the absence of tunneling, $\Delta = 0$. The eigenstates in this case can be written as
\begin{equation}
  \ket{\tau, \vect{n}} = e^{- i \tau^z S} \ket{\tau} \otimes \prod_k \frac{1}{\sqrt{n_k!}} \left(b_k^\dagger \right)^{n_k} \ket{0} = e^{- i \tau^z S} \ket{\tau} \otimes \ket{\vect{n}}
\end{equation}
where $\tau = m, m_\perp$ labels the two eigenstates of $\tau^z$. The unitary $S$ diagonalizes the system bath interaction,
\begin{equation}
  e^{i \tau^z S} \left\{ \sum_k \omega_k b_k^\dagger b_k + \tau^z \otimes \left[ \sum_k g_k \left( b_k + b_k^\dagger \right) + \sum_{k,l} \frac{g_k g_l}{E} \left(b_k + b_k^\dagger \right) \left( b_l + b_l^\dagger \right)  \right] \right\} e^{- i \tau^z S} = \sum_k \omega_k b_k^\dagger b_k,
\end{equation}
where we dropped all terms acting only on the system on the right-hand side of the equation. We have also neglected corrections to the spectrum of the environment by the system--environment interaction since they are inversely proportional to the volume of the bath, as already observed in the derivation of the effective Hamiltonian in section~\ref{app:effective_hamiltonian}. The Hamiltonian to be diagonalized here is in fact very similar to that in section~\ref{app:effective_hamiltonian} and the same methods can be applied. We obtain
\begin{gather}
  e^{- i \tau^z S} = e^{- i \tau^z S_1} e^{- i \tau^z S_2}
  \label{eq:s1s2}
\end{gather}
where
\begin{equation}
  S_1 = i \sum_k \delta_k \left( b_k - b_k^\dagger \right), \qquad 
  \delta_k =   \frac{1}{1-a^2} \frac{g_k}{\omega_k}, \qquad
  a = \frac{4}{E} \int \di \omega \, \frac{J(\omega)}{\omega}
\end{equation}
effects a displacement to remove the single-boson terms, while
\begin{equation}
  S_2 = \sum_{k,l} \left[ A_{kl} b_k^\dagger b_l + \frac{i}{2} B_{kl} \left( b_k b_l - b_k^\dagger b_l^\dagger \right) \right], \qquad
  A_{kl} = - \frac{2 i}{\omega_k - \omega_l} \frac{g_k g_l}{E} (1 - \delta_{kl}), \qquad
  B_{kl} = - \frac{2}{\omega_k + \omega_l} \frac{g_k g_l}{E}
\end{equation}
diagonalizes the two-boson terms in the weak coupling limit, $J(\omega) \ll E$. We point out that we omitted a term proportional to $\tau^z$ in the expression for the displacement $\delta_k$. Such a term merely gives rise to a state-independent displacement in Eq.~(\ref{eq:s1s2}), which does not affect the renormalized tunneling rate as will be apparent shortly. In addition, we will drop the pre-factor involving $a$ since we are only interested in the asymptotic scaling with the size of the search space.

For a given set of occupation number $\vect{n}$, the renormalized tunneling rate can now be expressed as
\begin{equation}
  \tilde \Delta_\vect{n} = \Delta \bra{m_\perp, \vect{n}} \tau^x \ket{m, \vect{n}}' = \Delta \bra{\vect{n}} e^{- i S_2} e^{- 2 i S_1} e^{- i S_2} \ket{\vect{n}}'.
  \label{eq:expectation_value}
\end{equation}
Here the prime reminds us that we should only consider processes that are fast compared to the dynamics of the system. At zero bias, the only time scale of the system is set by the renormalized tunneling rate. Hence, the renormalized tunneling rate may be determined self-conistently by evaluating the expectation value in Eq.~(\ref{eq:expectation_value}) with a low-frequency cutoff $\Omega = p \tilde \Delta_\vect{n}$, where $p$ is an unimportant numerical factor as long as $p \gg 1$. We point out this argument readily generalizes to the case of finite bias, where the cutoff should be taken to be $\Omega = p \sqrt{\varepsilon^2 + \tilde \Delta_\vect{n}^2}$. We do not discuss this more complicated case here since the transition between the coherent and incoherent regime first occurs at the smallest gap of the system, that is, at zero bias. As argued in the main text, any potential quantum speedup is lost when the system is rendered incoherent during any part of the evolution. Thus, the coherence properties at zero bias fully determine the performance of the algorithm.

The renormalized tunneling rate in Eq.~(\ref{eq:expectation_value}) clearly depends on the occupation numbers $\vect{n}$ and it is therefore not unique at finite temperature. Nevertheless, we can obtain a typical value $\tilde \Delta$ by taking a thermal expectation value
\begin{equation}
  \tilde \Delta = \Delta \tr \left\{ \rho e^{- i S_2} e^{- 2 i S_1} e^{- i S_2} \right\}',
  \label{eq:renormalized_final}
\end{equation}
where $\rho = e^{- \sum \omega_k b_k^\dagger b_k /T} / \mathcal{Z}$ is the thermal state at temperature $T$. 

\subsection{Single-boson processes}
Before we consider Eq.~(\ref{eq:renormalized_final}) fully, it is instructive to compute the renormalized tunneling rate in the absence of two-boson processes, i.e., setting $S_2 = 0$. The trace in Eq.~(\ref{eq:renormalized_final}) is most readily evaluated by observing that
\begin{equation}
  \tilde \Delta = \Delta \langle e^{- 2 i S_1} \rangle = \Delta \, e^{- 2 \langle S_1^2 \rangle},
\end{equation}
since $S_1$ is linear in the bosonic operators and the expectation value is with respect to a Gaussian state. Hence
\begin{equation}
  \tilde \Delta = \Delta \exp \left[ - 2 \sideset{}{'}\sum_k  \delta_k^2 (1 + 2 N(\omega_k)) \right] = \Delta \exp \left[ - 2 \int_{p \tilde \Delta}^\infty \di \omega \, \frac{J(\omega)}{\omega^2} \coth \frac{\omega}{2 T} \right],
  \label{eq:renormalized1}
\end{equation}
where we cut off the integral at $p \tilde \Delta$ in accordance with the prescription of adiabatic renormalization. At zero temperature, the integral is convergent for $\eta >1$ and we can safely extend the lower limit to $0$:
\begin{equation}
  \tilde \Delta = \Delta \exp \left[ - 2 \int_0^\infty \di \omega \, \frac{J(\omega)}{\omega^2} \right] \qquad \text{if } \eta > 1.
\end{equation}
When $\eta < 1$, the integral diverges with small $\tilde \Delta$ as $\tilde \Delta^{\eta - 1}$. Thus,
\begin{equation}
  \tilde \Delta \approx \Delta \exp \left[ - \frac{2 \alpha}{1 - \eta} \left( p \tilde \Delta \right)^{\eta - 1} + \log c \right],
\end{equation}
where the constant $c$ depends on the high frequency behavior of the noise spectrum $J(\omega)$. We re-arrange the expression to
\begin{equation}z
  \left( \frac{\tilde \Delta}{ c \Delta} \right)^{1 - \eta} \log \left( \frac{\tilde \Delta}{ c \Delta} \right) = - \frac{2 \alpha}{1 - \eta} \left( \frac{1}{p c \Delta} \right)^{1-\eta}.
\end{equation}
The expression on the left has a global minimum of $- 1/e (1- \eta)$ such that a non-zero solution for $\tilde \Delta$ only exists if
\begin{equation}
  2 \alpha \left( \frac{1}{p c \Delta} \right)^{1 - \eta} < \frac{1}{e},
\end{equation}
or
\begin{equation}
  \alpha < \frac{1}{2 e} \left( p c \Delta \right)^{1 - \eta}.
\end{equation}
This shows that there exists a critical coupling strength
\begin{equation}
  \alpha^* \propto \Delta^{1 - \eta} = O(N^{(\eta - 1)/2})
  \label{eq:critical_coupling}
\end{equation}
above which $\tilde \Delta = 0$. Our simple argument does not predict the precise value of $\alpha^*$ due to the dependence on $p$. Nevertheless, more detailed studies have confirmed that the form of Eq.~(\ref{eq:critical_coupling}) is qualitatively correct~\cite{Kehrein1996,Bulla2003}. This result implies that for a fixed $\alpha$, the dynamics are incoherent even at zero temperature for sub-ohmic environments in the limit of large $N$. 

At $\eta = 1$, the exponent in Eq.~(\ref{eq:renormalized1}) diverges logarithmically with $\tilde \Delta$ such that
\begin{equation}
  \tilde \Delta \propto \Delta^{1/(1 - 2 \alpha)} = O(N^{- 1/2(1 - 2 \alpha)}).
\end{equation}
This expression is only valid if $\tilde \Delta < \Delta$, which implies that $\alpha < 1/2$. If $\alpha > 1/2$, the renormalized tunneling rate vanishes. The critical coupling strength is therefore independent of $N$ at $\eta = 1$, as expected from Eq.~(\ref{eq:critical_coupling}).

The above arguments can be readily generalized to the case of finite temperature. Assuming that $T \gg \tilde \Delta$, as will be naturally the case for large systems, we can can approximate $\coth \omega/2T \approx 2T / \omega$ near the lower limit of the integral. The integral is convergent for $\eta > 2$ and we obtain to a good approximation
\begin{equation}
  \tilde \Delta = \Delta \exp \left[ - 2 \int_0^\infty \di \omega \, \frac{J(\omega)}{\omega^2} \coth \frac{\omega}{2T} \right] \qquad \text{if } \eta > 2.
\end{equation}
For $\eta >2$, adiabatic renormalization predicts that $\tilde \Delta = 0$ unless
\begin{equation}
  \alpha < \frac{1}{4 e T} \left( p c \Delta \right)^{2 - \eta}.
\end{equation}
For fixed temperature, there exists a critical coupling strength
\begin{equation}
  \alpha^* \propto \frac{\Delta^{2 - \eta}}{T} = O(N^{(\eta - 2)/2})
\end{equation}
above which $\tilde \Delta = 0$. For fixed $\alpha$, we can alternatively identify a critical coupling temperature with the same scaling as the critical coupling strength,
\begin{equation}
  T^* \propto \frac{\Delta^{2 - \eta}}{\alpha} = O(N^{(\eta - 2)/2}).
  \label{eq:critical_temp}
\end{equation}
The renormalized tunneling rate vanishes for any $T > T^*$. At $\eta = 2$, the renormalized tunneling rate vanishes unless $\alpha T < 1/4$, showing that the critical coupling strength and temperature are again independent of the search space size.

We point out that Eq.~(\ref{eq:critical_temp}) is only valid if $\eta > 1$ since otherwise the assumption that $T \gg \tilde \Delta$ cannot be satisfied in the limit of large $N$. For $\eta < 1$, we found above that the incoherent tunneling rate vanishes even at zero temperature in the limit of large $N$ and fixed $\alpha$. This can be summarized as
\begin{equation}
  T^* = 
  \begin{cases}
    0, & \text{if }\eta < 1\\
    O(N^{(\eta-2)/2}), & \text{if } \eta > 1
  \end{cases}.
\end{equation}
At $\eta = 1$, a non-zero critical temperature, which scales as $T^* = O(N^{-1/2})$ only exists if $\alpha < 1/2$.

\subsection{Two-boson processes\label{app:two_boson}}
We repeat the above analysis for two-boson processes, ignoring single boson processes for the moment. The expectation value in Eq.~(\ref{eq:renormalized_final}) is harder to compute in this case since $S_2$ is a quadratic operator involving terms of the form $b_k^\dagger b_l$, $b_k b_l$, and $b_k^\dagger b_l^\dagger$. We will only evaluate it perturbatively by expanding the exponential to the lowest non-trivial order. By ignoring terms that scale inversely with the volume of the bath, we  hence obtain
\begin{align}
  \tilde \Delta &\approx \Delta \exp \left[ - 2 \sideset{}{'} \sum_{k,l} A_{kl} A_{lk} N(\omega_k) \left( 1 + N(\omega_l) \right) - \sideset{}{'} \sum_{k,l} B_{kl} B_{lk} \left[ \left( 1 + N(\omega_k) \right) \left( 1 + N(\omega_l) \right) - N(\omega_k) N(\omega_l)  \right] \right]\\
  &\approx \Delta \exp \left[- \phi - \chi \right],
\end{align}
where
\begin{align}
  \phi &= 2 \int' \di \omega \, \di \omega' \, \frac{J(\omega) J(\omega')}{(\omega - \omega')^2} N(\omega) \left( 1 + N(\omega') \right) ,\label{eq:phi}\\
  \chi &= \int' \di \omega \, \di \omega' \, \frac{J(\omega) J(\omega')}{(\omega + \omega')^2} \left[ ( 1 + N(\omega))(1 + N(\omega')) - N(\omega) N(\omega') \right] .
\end{align}
In both cases, the integral is to be taken over processes that are fast compared to the low-frequency cutoff $p \tilde \Delta$. For $\phi$, which describes two-boson scattering processes, this corresponds to $|\omega - \omega'| > p \tilde \Delta$, stating that the beating frequency of the two modes is fast. The two-boson absorption and emission processes are captured by $\chi$, for which we therefore impose that $\omega + \omega' > p \tilde \Delta$. We point out that the above expansion is only justified if $A_{kl}$ and $B_{kl}$ as well as $\sqrt{N(\omega_k) N(\omega_l)} A_{kl}$ and $\sqrt{N(\omega_k) N(\omega_l)} B_{kl}$ are small matrices in the sense that each column forms a vector with magnitude much less than one. This is indeed the case for $A_{kl}$ and $B_{kl}$ in the weak coupling limit $J(\omega) \ll E$, as shown explicitly in section~\ref{app:effective_hamiltonian}. A similar treatment can be applied to the other two matrices, giving rise to the condition
\begin{equation}
  J(\omega) N(\omega) \ll E.
  \label{eq:weak_coupling}
\end{equation}
At high frequencies $\omega \gg T$, this is trivially satisfied in the weak coupling limit. At low frequencies $\omega \ll T$, however, this leads to the additional constraint
\begin{equation}
  \frac{J(\omega)}{\omega} \ll \frac{E}{T}.
\end{equation}
It is important to to note that this inequality can only be satisfied as $\omega \to 0$ for ohmic and super-ohmic environments. Restricting ourselves to this particular parameter regime is not a significant limitation since single-boson processes alone will render the dynamics incoherent even at zero temperature for sub-ohmic environments. Therefore, two-boson processes are expected to modify the dynamics qualitatively only for ohmic and super-ohmic environments.

We now investigate the low-frequency divergences of $\phi$ and $\chi$ to identify the critical coupling strength and temperature as in the case of single-boson processes. At zero temperature, $\phi$ vanishes while $\chi$ is always finite. Hence, two-boson processes only weakly modify the tunneling rate at zero temperature. At non-zero temperatures, $\chi$ remains finite whereas $\phi$ exhibits an infrared divergence for any $\eta$. The functional form of the divergence with the low-frequency cutoff $p \tilde \Delta$ can be found to be given by
\begin{align}
  \phi \propto \frac{\alpha^2 T^{2 \eta + 1}}{E^2 \tilde \Delta}.
\end{align}
Using similar arguments to the ones for the single-boson processes, this allows us to identify a critical coupling strength
\begin{equation}
  \alpha^* \propto E T^{-(\eta + 1/2)} \tilde \Delta^{1/2} = O(N^{-1/4})
\end{equation}
and a critical temperature
\begin{equation}
  T^* \propto \alpha^{-2/(2 \eta + 1)} E^{2/(2 \eta + 1)} \tilde \Delta^{1/(2 \eta + 1)} = O(N^{-1/(4 \eta + 2)}).
\end{equation}

The divergence of $\phi$ as $\tilde \Delta \to 0$ originates from the denominator in Eq.~(\ref{eq:phi}), the form of which is dictated by conservation of energy during the scattering process. It is for this reason that higher-order terms in the effective Hamitonian Eq.~(\ref{eq:heff}) are not expected to modify the scaling of the critical temperature or the critical coupling strength.

\subsection{Combined effects}
We will now show that the single-boson and two-boson processes approximately decouple in the regime of interest. We start by expressing Eq.~(\ref{eq:renormalized_final}) in a coherent state basis
\begin{equation}
  \tilde \Delta = \Delta \int \Di \alpha \, \Di \beta \, \bra{\alpha} e^{- i S_2} \rho e^{- i S_2} \ket{\beta} \bra{\beta} e^{ - 2 i S_1} \ket{\alpha},
  \label{eq:coherent_states}
\end{equation}
where we introduced the short-hand notation $\Di \alpha = \prod_k \di^2 \alpha_k / \pi$ and $\ket{\alpha} = \prod_k \ket{\alpha_k}$. The state $\ket{\alpha_k}$ is a coherent state of the mode $b_k$ and the integral runs over the entire complex plane for each mode. We note that the matrix elements may be written as 
\begin{align}
  \bra{\alpha} e^{- i S_2} \rho e^{- i S_2} \ket{\beta} &= e^{-f(\alpha^*, \beta)} \inner{\alpha}{\beta},\\
  \bra{\beta} e^{ - 2 i S_1} \ket{\alpha} &= e^{g(\beta^*, \alpha)} \inner{\beta}{\alpha},
\end{align}
where $g(\beta^*, \alpha)$ is a linear function, while $f(\alpha^*, \beta)$ contains only quadratic terms \cite{Gardiner2004}. Computing the function $f$ is rather cumbersome due to the presence of the squeezing terms $b_k b_l$ and $b_k^\dagger b_l^\dagger$ in $S_2$. For our purposes, it suffices to exploit the general structure of a Gaussian integral over a real vector $\vect{v}$,
\begin{equation}
  \int \left( \prod_n \frac{\di v_n}{\sqrt{2 \pi}} \right) \exp \left[- \frac{1}{2} \vect{v}^T M \vect{v} + \vect{w}^T \vect{v} \right] = \frac{1}{\det M} \exp \left[ \vect{w}^T M^{-1} \vect{w} \right].
\end{equation}
Applied to Eq.~(\ref{eq:coherent_states}), we can see that the matrix $M$ is determined by $S_2$, while the vector $\vect{w}$ follows from $S_1$. We may thus write
\begin{equation}
  \tilde \Delta = \Delta \left\langle e^{- 2 i S_2^{\phantom{'}}} \right\rangle \left\langle e^{- 2 i S_1'} \right\rangle,
\end{equation}
where the operator $S_1'$ accounts for both single-boson processes and the coupling between single-boson and two-boson processes. Under the same conditions that we were able to expand $e^{- 2 i S_2}$ in the previous section, we can also expand $S_1'$ in powers of $A_{kl}$ and $B_{kl}$. To leading order, we clearly must have $S_1' \approx S_1$, which shows that the single-boson and two-boson processes decouple under the assumption that $J(\omega) \ll E$ and $J(\omega) N(\omega) \ll E$. The nature of the dynamics of the system may thus be deduced by considering the two processes separately, as done in the main text.

\section{Thermalization rates\label{app:thermalization}}
\subsection{Incoherent regime}
We argued in the main text that it is necessary to determine the scaling of the thermalization rate in order to exclude the possibility of a quantum speedup in the incoherent regime. In particular, a speedup over the classical algorithm is possible if the thermalization rate decays slower with the size of the search space than $N^{-1}$. In the incoherent regime, adiabatic renormalization predicts that the system is localized. However, adiabatic renormalization does not take into account incoherent tunneling. To estimate the rate of incoherent tunneling, we perform perturbation theory in the bare tunneling rate $\Delta$. It is convenient to switch to the basis that diagonalizes the Hamiltonian in the absence of tunneling, 
\begin{equation}
  e^{iS} H e^{- i S} = - \frac{1}{2} \left( \varepsilon \tau^z + \Delta e^{i S} \tau^x e^{- i S} \right) + \sum_k \omega_k b_k^\dagger b_k,
\end{equation}
where $S$ is given by Eq.~(\ref{eq:s1s2}). We move to the interaction picture, where the time-evolution is fully governed by
\begin{equation}
  V(t) = - \frac{\Delta}{2} e^{- i \varepsilon t} e^{i S(t)} \tau^+ e^{- i S(t)} + \text{h.c.},
\end{equation}
with $\tau^+ = (\tau^x + i \tau^y)/2$ and
\begin{equation}
  S(t) = e^{i t \sum_k \omega_k b_k^\dagger b_k} S e^{- i t \sum_k \omega_k b_k^\dagger b_k}.
\end{equation}
Starting with the initial state $\ket{\psi(0)} = \ket{m_\perp, \vect{n}}$, the probability that the system ends up in $\ket{m}$ after time $t$ is given to lowest order in perturbation theory by
\begin{equation}
  p(t) = \sum_\vect{n'} \left| \bra{m, \vect{n'}} \int_0^t \di t' V(t') \ket{m_\perp, \vect{n}} \right|^2 = \int_0^t \di t' \int_0^t \di t'' \bra{m_\perp, \vect{n}} V(t') \ket{m}\bra{m} V(t'') \ket{m_\perp, \vect{n}}.
\end{equation}
By expressing $e^{iS}$ in terms of $S_1$ and $S_2$ and taking a thermal average over the initial state, we obtain
\begin{equation} 
  p(t) = \left( \frac{\Delta}{2} \right)^2 \int_0^t \di t' \int_0^t \di t'' e^{i \varepsilon (t' - t'')} \left \langle e^{i S_2(t')} e^{2 i S_1(t')} e^{i S_2(t')} e^{- i S_2(t'')} e^{- 2 i S_1(t'')} e^{- i S_2(t'')} \right\rangle
\end{equation}
We note that the expectation value in the integrand is a function of $t' - t''$ only, which allows us to write
\begin{equation}
  p(t) = \int_0^t \di t' \, \Gamma(t'), 
\end{equation}
where
\begin{equation}
  \Gamma(t) = \left( \frac{\Delta}{2} \right)^2 \int_{-t}^t \di t' \, e^{i \varepsilon t'} \left \langle e^{i S_2(t')} e^{2 i S_1(t')} e^{i S_2(t')} e^{- i S_2(0)} e^{- 2 i S_1(0)} e^{- i S_2(0)} \right\rangle
\end{equation}
is the instantaneous decay rate at time $t$. Typically, we can extend the limits of this integral to infinity to obtain a single decay rate
\begin{equation}
  \Gamma = \left( \frac{\Delta}{2} \right)^2 \int_{-\infty}^\infty \di t' \, e^{i \varepsilon t'} \left \langle e^{i S_2(t')} e^{2 i S_1(t')} e^{i S_2(t')} e^{- i S_2(0)} e^{- 2 i S_1(0)} e^{- i S_2(0)} \right\rangle.
\end{equation}
This is a good approximation for almost all $t$ provided the width over which the integrand contributes significantly is small compared to $1/\Gamma$. Since the width of the integrand is independent of $N$, the decay rate scales as $O(N^{-1})$, and thus the condition for extending the limits of the integral is always fulfilled in the limit of large $N$. We will not evaluate the above expression any further as we are only interested in the scaling with the search space size.

\subsection{Coherent regime}
For completeness we briefly discuss the thermalization rate in the coherent regime, i.e., $\eta > 1$ at zero temperature. The thermalization rate in this regime has no immediate implications for the scalability of the quantum algorithm since a quantum speedup is always available in the coherent regime. However, a thermalization rate that exceeds the classical scaling $O(N^{-1})$ enables a quantum speedup by thermalization alone as discussed in the main text.

In the coherent regime, thermalization occurs via transitions between the eigenstates of the closed system rather than by incoherent tunneling. The thermalization rate is readily obtained by applying Fermi's Golden rule after adiabatic renormalization. For the single-boson processes, this yields at zero bias
\begin{equation}
  \Gamma_1 = 2 \pi \sum_k g_k^2 \delta(\omega_k - \tilde \Delta)  = 2 \pi J(\tilde \Delta) = O(N^{- \eta/2}) .
\end{equation}
Interestingly, the thermalization rate exceeds the classical limit for $1 < \eta < 2$. We further remark that the rate drops below the classical scaling for $\eta > 2$. In this regime, incoherent tunneling and processes coupling to $\tau^x$ and $\tau^y$, which we have neglected, will dominate the thermalization rate. 

For the the two-boson processes, the Golden rule rate is given by
\begin{equation}
  \Gamma_2 = \frac{2 \pi}{E^2} \int_0^{\tilde \Delta} \di \omega \, J(\omega) J(\tilde \Delta - \omega) = O(N^{-(\eta + 1/2)}).
\end{equation}
This vanishes parametrically faster than the single-boson decay rate such that two-boson emission only contributes weakly to the thermalization rate.

\section{Diagonalization of quadratic Hamiltonians\label{app:diagonalization}}
We briefly review the diagonalization of a general quadratic Hamiltonian of the form
\begin{equation}
  H = \beta^\dagger M \beta ,
  \label{eq:quadratic_hamiltonian}
\end{equation}
where $M$ is a Hermitian matrix and $\beta = (b_1, b_2, \dots, b_1^\dagger, b_2^\dagger, \dots)$ is a vector formed by creation and annihilation operators. We closely follow the notation of reference~\cite{Blaizot1986}, where the diagonalization of both fermionic and bosonic Hamiltonians is discussed in detail. For the sake of clarity, we focus on bosons below.

The goal is to introduce new bosonic operators $\gamma = (c_1, c_2, . . . , c_1^\dagger, c_2^\dagger, . . .)$ such that the Hamiltonian can be written as
\begin{equation}
  H = \sum_i \lambda_i c_i^\dagger c_i
  \label{eq:diagonal_hamiltonian}
\end{equation}
up to a constant. The new operators $\gamma$ are related to the original operators by a linear transformation
\begin{equation}
  \begin{pmatrix}
    c\\
    c^\dagger
  \end{pmatrix}
  = T
  \begin{pmatrix}
    b\\
    b^\dagger
  \end{pmatrix}.
  \label{eq:linear_transformation}
\end{equation}
The fact that $b^\dagger$ is the adjoint of $b$ implies that $T$ must take the form
\begin{equation}
  T = 
  \begin{pmatrix}
    A & B\\
    B^* & A^*
  \end{pmatrix}.
  \label{eq:constraint_t1}
\end{equation}
Furthermore, the conservation of canonical commutation relations leads to the additional constraint
\begin{equation}
  T^{-1} = \mu T^\dagger \mu, \qquad
  \mu = 
  \begin{pmatrix}
    I & 0\\
    0 & -I
  \end{pmatrix}.
  \label{eq:constraint_t2}
\end{equation}
Eq.~(\ref{eq:quadratic_hamiltonian}) can hence be written as
\begin{equation}
  H = \gamma^\dagger \mu T \mu M T^{-1} \gamma .
  \label{eq:transformed}
\end{equation}
Assuming that $M$ is positive definite, it is shown in~\cite{Blaizot1986} that there exists a transformation $T$ satisfying Eq.~(\ref{eq:constraint_t1}) and Eq.~(\ref{eq:constraint_t2}) which diagonalizes $\mu M$ to give
\begin{equation}
  T \mu M T^{-1} = \frac{1}{2}
  \begin{pmatrix}
    \lambda_i \delta_{ij} & 0\\
    0 & - \lambda_i \delta_{ij}
  \end{pmatrix}.
\end{equation}
This immediately yields the desired result Eq.~(\ref{eq:diagonal_hamiltonian}) up to a constant.

It is often useful to express the transformation described by the matrix $T$ as a unitary transformation $S$ acting on the creation and annihilation operators, i.e.~
\begin{equation}
  c_i = S b_i S^\dagger, \qquad S^\dagger S = S S^\dagger = I.
  \label{eq:unitary}
\end{equation}
The transformation takes the form $S = \exp \left( i \beta^\dagger K \beta /2 \right)$, where $K$ is a Hermitian matrix. By direct substitution into Eq.~(\ref{eq:unitary}) and comparison to Eq.~(\ref{eq:linear_transformation}) we obtain
\begin{equation}
  T = e^{- i \mu K}.
\end{equation}

Finally, we note that the vacuum $\ket{0_c}$, where $c_i \ket{0_c} = 0$ for all $i$, is related to the vacuum $\ket{0_b}$, for which $b_i \ket{0_b} = 0$, by
\begin{equation}
  \ket{0_c} = S \ket{0_b}.
\end{equation}
All other Fock states transform in the same manner.

\bibliography{bibliography}